\documentclass[aip,reprint]{revtex4-1}

\usepackage[english]{babel}
\usepackage[latin1]{inputenc}
\usepackage{amssymb}
\usepackage{mathrsfs}
\usepackage{amsmath}
\usepackage[dvips]{graphicx}
\usepackage{epsfig,bm}
\usepackage{graphicx}
\usepackage{amsfonts}
\usepackage{graphics}
\usepackage{color}
\usepackage{setspace}
\usepackage{amscd}
\usepackage{amsxtra}
\usepackage{amstext}
\usepackage{latexsym}
\usepackage{dcolumn}
\usepackage{bm}
\usepackage{graphics}
\usepackage{url}

\draft 

\begin{document}


\title{Dynamic Range in the C.elegans Brain Network} 



\author{Chris G. Antonopoulos}
\email[]{canton@essex.ac.uk}
\affiliation{Department of Mathematical Sciences, University of Essex, Wivenhoe Park, CO4 3SQ Colchester, UK}


\date{\today}

\begin{abstract}
We study external, electrical perturbations and their responses in the brain dynamic network of the \textit{Caenorhabditis elegans} soil worm, given by the connectome of its large somatic nervous system. Our analysis is inspired by a realistic experiment where one stimulates externally specific parts of the brain and studies the persistent neural activity triggered in other cortical regions. In this work, we perturb groups of neurons that form communities, identified by the walktrap community detection method, by trains of stereotypical electrical Poissonian impulses and study the propagation of neural activity to other communities by measuring the corresponding dynamic ranges and Steven law exponents. We show that when one perturbs specific communities, keeping the rest unperturbed, the external stimulations are able to propagate to some of them but not to all. There are also perturbations that do not trigger any response. We found that this depends on the initially perturbed community. Finally, we relate our findings for the former cases with low neural synchronization, self-criticality and large information flow capacity, and interpret them as the ability of the brain network to respond to external perturbations when it works at criticality and its information flow capacity becomes maximal.
\end{abstract}


\maketitle 


\textbf{One of the fundamental findings in neuroscience is the modular organization of the brain, which in turn points to an inherent parallel nature of brain computations. Cortical networks are hierarchical and clustered with complex connectivity patterns. Modular processors have to be sufficiently isolated and dynamically differentiated to achieve independent computations, but also globally connected to be integrated in coherent functions. One of the oldest problems in psychophysics is the relation between external stimulus and neural response in such brains. This is a nonlinear relation as a linear one would lead to an unbounded neural response, not observed in real sensorial organs and, can be quantified by means of the dynamic range theory. In this paper, we explore the effect of the application of external electrical stimuli, modelled by Poisson processes of fixed average input rates, to specific communities of the brain dynamical network of the \textit{Caenorhabditis elegans} soil worm, and study external perturbations spreading to the rest by measuring the dynamic ranges and Steven law exponents. Our work is inspired by a realistic experiment where one stimulates externally specific cortical regions and studies the persistent neural activity triggered in other regions. We show that when particular communities are perturbed, keeping the rest unperturbed, the external stimulations propagate to some but not all of them. In such cases, the dynamic ranges become maximal. We interpret our results as the ability of the brain network to respond to external stimulations when it works at self-criticality and its information flow capacity becomes maximal.}

\section{Introduction}

Many phenomena in nature can be described by networks. Neuroscientists have used tools for the analysis of complex networks that help realize the functionality and structure of the brain. It was found that many aspects of brain network structures can be found in a wide range of non-neural complex networks \cite{Meunieretal2001,Yamaguti2014} as well. One of the main findings in neuroscience is the modular organization of the brain, which in turn implies an inherent parallel nature of brain computations \cite{Meunieretal2001}. Modular processors have to be sufficiently isolated and dynamically differentiated to achieve independent computations, but also globally connected to be integrated in coherent functions \cite{Meunieretal2001}. Moreover, it has been found that the cortical network is hierarchical and clustered with a complex connectivity \cite{Hilgetag_2004} pattern. A possible network description for this modular organization is that brain networks may be small-world structured \cite{Heetal2007} with properties similar to other complex non-neural networks \cite{Stam2004}. This idea is further supported by the systematic finding of small-world topology in a wide range of human brain networks derived from structural \cite{Heetal2007}, functional \cite{Eguiluzetal2005}, and diffusion tensor MRI \cite{Hagmannetal2007} studies. The small-world topology has also been identified at the cellular-network scale in functional cortical neural circuits in mammals \cite{Yuetal2008} and also in the nervous system of the nematode \textit{Caenorhabditis elegans} (\textit{C.elegans}) \cite{Wattsetal1998}.

One of the oldest problems in psychophysics is the relation between external stimulus and neural response \cite{Stevens1975}. It is expected to be nonlinear as a linear one would lead to an unbounded neural response, not observed in real sensorial organs. In contrast, neurons actually exhibit a limited ability to respond to external stimulations. It is thus reasonable to consider a kind of saturation during which the response remains practically the same as the stimulus continues to increase. Stevens \cite{Stevens1975} proposed that this relation is given by the power-law $P\sim I^m$, where $P$ is the magnitude of the response, $I$ the magnitude of the applied stimulus and $m$ a positive response exponent that is usually smaller than 2. $m$ depends on the type of stimulus and attains different values. For  example, it was found that the response to a vibrating plate captured by the sensorial organs in the finger leads to $m=0.95$ and $m=0.6$ for 60 Hz and 250 Hz vibrations, respectively.

This power-law dependence is valid within some bounds for $I$ due to anatomical and physiological limitations. To quantify the response intensity within these bounds, one can employ the dynamic range which is the difference between the smallest and largest response intensity, expressed in decibels (dB). According to \cite{Siegeletal1982}, the human senses of sight and hearing have a large dynamic range of 90 dB and 100 dB, respectively. The variability of the values of $m$ and the big differences in the dynamic range for different sensory organs are currently interesting subjects in computational neuroscience. In this context, it is interesting to investigate macroscopically observed features such as the dynamic range and exponent $m$ driven by the application of external, locally applied, electrical perturbations affecting the underlying complex neuronal dynamics. By locally, we mean the application of external stimulations to particular ensembles or groups of neurons of the brain network. In such cases, the firing rate of a neuron $F$, i.e. the number of spikes per time unit, can be used to study its response to external stimuli. Initially, the external stimulus is injected to a group of neurons as an input current. Usually, this has the form of a sequence of random stereotypical electrical impulses driven by a constant average input rate $r$. One then studies macroscopically the stimulus-response curve for single neurons or ensembles of neurons that perform similar tasks. This curve provides the input rate $r$ - firing rate $F$ relation in the form of a power-law dependence within some bounds for $r$, for which the curve does not saturate but increases (see for example Fig. \ref{fig:average_firing_rate_p}). Consequently, one then estimates the exponent $m$ of the power-law dependence by a linear fitting, and the dynamic range $\Delta$ can be estimated by the bounds of the input rate $r$ for which $m$ is estimated.

There is experimental evidence suggesting that the dynamic range of a single neuron is smaller than the dynamic range macroscopically observed for ensembles of neurons. For example, the dynamic range of neurons of the olfactory system is about 10 dB whereas the corresponding dendro-dendritic neural network in the glomeruli has three times larger dynamic range \cite{Rospars2000133,Rosparsetal2003}. The reason is that $\Delta$ can be thought of as a collective effect attributed to the topology of the network. There are two mechanisms that contribute to the dynamic range and makes it larger for ensembles of neurons. The first one is the intrinsic threshold variations in networks of sensory neurons \cite{Clelandetal1999} and the second the adaptation of single neurons to the statistics of the ambient stimuli \cite{Kim15022003}. It has been recently found in \cite{Deans2002703} that the elimination of gap junctions attributed to electrical synapses causes a decrease of the dynamic range and an increase of the exponent $m$. Thus, the enhancement of the dynamic range is possible by considering simultaneously electrical and chemical connections between different neurons. The authors in \cite{2014PhyA..410..628B} studied the dynamic range in small-world networks of Hodgkin-Huxley neurons with chemical synapses and found that the dynamic range of sensory organs is larger than that of single neurons and that it increases with the network size. They suggested that the enhancement of the dynamic range observed in sensory organs, with respect to single neurons, is an emergent property of complex networks dynamics. However, their study does not take into consideration electrical synapses for local neural connectivity, as the authors in Ref. \cite{Viana2014164} do. Even though most of the neural synapses involved in sensory organs are electrical (i.e. gap junctions), chemical synapses are also important to take into consideration in more realistic models.

In this work, we go one step further and study, by means of the dynamic range theory, the effect of the application of external electrical stimuli, modelled by Poisson processes of fixed average input rate $r$, to specific neural ensembles (i.e. communities) of the brain dynamical network (BDN) of the \textit{C.elegans} soil worm (see Subsec. \ref{subSubsection_C.elegans_data} of the Methodology for the details), considering at the same time electrical and chemical connections. We aim to study external perturbations spreading to other communities of the BDN. By BDN we mean a brain network, given here by the connectome of the \textit{C.elegans} brain, in which each node is equipped with neural dynamics (see Subsec. \ref{Subsection_Hindmarsh-Rose_Model_for_Brain_Dynamics}). The novelty of this work is two-folded: First, we use the \textit{C.elegans} BDN as it is a realistic small-world network \cite{Antonopoulosetal2015}, almost completely mapped \cite{Varshneyetal2011}. We implement the external perturbations as Poisson processes because they are characterized by the property that the probability of firing a spike is independent of the firing activity at previous times. This implies that the interspike intervals $\tau$ are independent of the past spiking activity. The second novelty is that we excite single communities of the \textit{C.elegans} BDN independently by supplying external electrical stimuli to all neurons of the perturbed community, following a Poisson process, while keeping the rest in an unperturbed state. We use the same communities identified in Ref. \cite{Antonopoulosetal2015} in the \textit{C.elegans} BDN (see Subsec. \ref{Subsection_analysis_of_communities} of Methodology). We endow each neuron of the BDN with Hindmarsh-Rose dynamics, represented by Eq. \eqref{HR_model_1neuron}, and couple them locally, within each community, with electrical connections of strength $g_l$ and nonlocally (i.e. intercommunity communication) with chemical connections of strength $g_n$ (for the details see Subsec. \ref{Subsection_Hindmarsh-Rose_Model_for_Brain_Dynamics}). This setup, although artificial, is reminiscent of the modular organization of the brain and allows us to model the neural activity of the BDN and extract useful conclusions for the response of neural ensembles to external stimuli applied in different parts. In our study, we prepare parameter spaces for different coupling strengths $g_n$ and $g_l$ for the chemical and electrical connections, respectively, and, compute the dynamic range $\Delta$ and Steven law exponent $m$ for the different communities and the full network. This approach allows us to understand for which coupling regions, the external electrical perturbations $I$ applied to particular communities of the \textit{C.elegans} spread to others and to the full BDN. We do that by testing the power-law dependence of the stimulus-response relation within some bounds of the average input rate of the Poisson process and the ranges of chemical and electrical coupling strengths for which an enhancement of the dynamic range can be observed.

The paper is organized as follows: In Sec. \ref{section_online_methods} we outline the details and methodology of our study and in Sec. \ref{section_results} we present our results for the dynamic range, for external perturbations applied to different communities of the \textit{C.elegans} brain network. Finally, in Sec. \ref{section_conclusions} we discuss about our results and possible relations with neural synchronization, self-criticality and information flow in BDNs.

\section{Methodology}\label{section_online_methods}

\subsection{\textit{C.elegans} Data}\label{subSubsection_C.elegans_data}

We base our analysis for the BDN of the \textit{C.elegans} soil worm on Ref. \cite{Antonopoulosetal2015}. Particularly, we use the connectome of the large somatic nervous system found in Ref. \cite{cmtkdataset} that consists of 277 neurons. We use the undirected version of the adjacency matrix provided there because we are not concerned here with the direction of the information flow, and simulate the dynamics of the single neurons by HR neural dynamics given in Eq. \eqref{HR_model_1neuron}. We couple them by the corresponding adjacency matrix obtained from the connectome of the \textit{C.elegans} using Eqs. \eqref{HR_model_Nneurons}.

\subsection{Hindmarsh-Rose Model for Brain Dynamics}\label{Subsection_Hindmarsh-Rose_Model_for_Brain_Dynamics}

The complexity of the circuitry of the nervous system of the \textit{C.elegans} is rather big, let alone the complexity of the human brain which contains about 86 billion neurons  and thousands times more synapses \cite{Azevedoetal2009}! We study the \textit{C.elegans} nervous system in order to develop understanding about the human nervous system and its response to external stimulations. This is because both humans' and \textit{C.elegans}' nervous systems consist of neurons and the communication or flow of information is passing through synapses that use neurotransmitters to perform brain activity. Many of these neurotransmitters are common in humans and in the \textit{C.elegans}, such as Glutamate, GABA, Acetylcholine and Dopamine.

A synapse is a junction between pairs of neurons and is a mean through which neurons communicate with each other. There are electrical and chemical synapses: An electrical synapse (gap junction) is a physical connection between two neurons which allows electrons to pass through neurons by a very small gap between nerve cells. Electrical synapses are bidirectional and of a local character, existing between neurons whose cells are close. They are believed to contribute to the regulation of synchronization in brain networks. In contrast, chemical synapses are special junctions through which the axon of the pre-synaptic neuron comes close to the post-synaptic cell membrane of another neuron or non-neural cell and chemical neurotransmitters are released. We use in our model both kinds of synapses to study a more realistic case but avoid to use directed chemical links as their nature is not yet perfectly identified \cite{Varshneyetal2011}.

Following Refs. \cite{Baptistaetal2010,Antonopoulosetal2015}, we endow the nodes of the \textit{C.elegans} brain network with Hindmarsh-Rose dynamics \cite{Hindmarshetal1984}
\begin{eqnarray}\label{HR_model_1neuron}
 \dot{p}&=&q-ap^3+bp^2-n+I^{\mbox{ext}}\nonumber,\\
 \dot{q}&=&c-dp^2-q\nonumber,\\
 \dot{n}&=&e[s(p-p_0)-n],
 \end{eqnarray}
where $p$ is the membrane potential, $q$ the fast current, $Na^{+}$ or $K^{+}$, and $n$ the slow current, for example $Ca^{2+}$. The rest of the parameters are defined as $a=1$, $b=3$, $c=1$, $d=5$, $s=4$ and $p_0=-1.6$. $I^{\mbox{ext}}$ is the external electrical stimulation that is applied to the neuron and can be either fixed during the application or a time-dependent function. For example, for $I^{\mbox{ext}}=3.25$, the system exhibits a multi-scale chaotic behavior termed spike bursting \cite{Hindmarshetal1984} (see for example panel (b) of Fig. \ref{fig:histogram}). Parameter $e$ modulates the slow dynamics of the system and was set to 0.005 so that each neuron is chaotic. For these parameters, the HR model enables the spiking-bursting behavior of the membrane potential observed in experiments made with a single neuron \textit{in vitro}.

We couple the HR system to create an undirected BDN of $N_n$ neurons connected simultaneously by electrical (linear diffusive coupling) and chemical (nonlinear coupling) links
\begin{widetext}
\begin{eqnarray}\label{HR_model_Nneurons}
 \dot{p}_i&=&q_i-a p_i^3+bp_i^2-n_i+I^{\mbox{ext}}_i-g_n(p_i-V_{\mbox{syn}})\sum_{j=1}^{N_n}\mathbf{B}_{ij}S(p_j)-g_l\sum_{j=1}^{N_n}\mathbf{G}_{ij}H(p_j)\nonumber,\\
 \dot{q}_i&=&c-dp_i^2-q_i\nonumber,\\
 \dot{n}_i&=&e[s(p_i-p_0)-n_i],\nonumber\\
 \dot{\phi}_i&=&\frac{\dot{q}_i p_i-\dot{p}_i q_i}{p_i^2+q_i^2},\;i=1,\ldots,N_n,
\end{eqnarray}
\end{widetext}
where $\dot{\phi}_i$ is the instantaneous angular frequency of the $i$-th neuron and $\phi_i$ its corresponding phase. We consider $H(p_i)=p_i$ and
\begin{equation}
 S(p_j)=\frac{1}{1+e^{-\lambda(p_j-\theta_{\mbox{syn}})}},
\end{equation}
with $\theta_{\mbox{syn}}=-0.25$, $\lambda=10$, and $V_{\mbox{syn}}=2$ to create excitatory BDNs, meaning that when the presynaptic neuron fires, it induces a firing activity to the postsynaptic neuron. In Eqs. \eqref{HR_model_Nneurons}, $g_n$ is the coupling strength associated to the chemical and $g_l$ to the electrical synapses. For these parameters and excitatory networks, $|p_i|<2$ and $(p_i-V_{\mbox{syn}})$ is negative. In this work we have only used excitatory connections between the coupled neurons. We use as initial conditions for each neuron $i$: $p_i=-1.30784489+\eta^r_i$, $q_i=-7.32183132+\eta^r_i$, $n_i=3.35299859+\eta^r_i$ and $\phi_i=0$, where $\eta^r_i$ is a uniformly distributed random number in $[0,0.5]$ for all $i=1,\ldots,N_n$, following Ref. \cite{Antonopoulosetal2015}. In Eq. \eqref{HR_model_Nneurons}, $I^{\mbox{ext}}_i$ is the external time-varying electrical perturbation applied to the $i$-th neuron and has the form of a Poisson process \eqref{poisson_process} of fixed average input rate $r$.

$\mathbf{G}_{ij}$ accounts for the way neurons are electrically (diffusively) coupled and is a Laplacian matrix (i.e. $\mathbf{G}_{ij}=\mathbf{K}_{ij}-\mathbf{A}_{ij}$, where $\mathbf{A}$ is the adjacency matrix of the electrical connections and $\mathbf{K}$ is the degree identity matrix based on $\mathbf{A}$), and so $\sum_{j=1}^{N_n}\mathbf{G}_{ij}=0$. $\mathbf{B}_{ij}$ is an adjacency matrix and describes how neurons are chemically connected and therefore its diagonal elements are equal to 0, giving thus $\sum_{j=1}^{N_n}\mathbf{B}_{ij}=k_i$, where $k_i$ is the degree of the $i$-th neuron. A positive off-diagonal value in both matrices in row $i$ and column $j$ means that neuron $i$ perturbs neuron $j$ with an intensity given by $g_l\mathbf{G}_{ij}$ (electrical diffusive coupling) or $g_n\mathbf{B}_{ij}$ (chemical excitatory coupling). Therefore, the adjacency matrix $\mathbf{C}$ of the \textit{C.elegans} brain network considered in this work is given by
\begin{equation}\nonumber
\mathbf{C}=\mathbf{A}+\mathbf{B}.
\end{equation}

\subsection{Numerical Simulations Details}\label{Subsection_numerical_simulations_details}

We integrate numerically Eqs. \eqref{HR_model_Nneurons} using the Euler integration method (order one) with time step $\delta t=0.01$. This allows us to reduce the numerical complexity and CPU time of the required simulations to feasible levels as a preliminary comparison of trajectories computed for the same parameters (i.e. $\delta t$, initial conditions, etc.) with integration methods of order 2, 3 and 4 revealed similar results. The numerical integration of the HR system of Eqs. \eqref{HR_model_Nneurons} was performed for the final integration time $t_f=5000$ and the computations of the different quantities start after the transient time $t_t=300$ to make sure that orbits converged to the attractor of the dynamics. Each point in the parameter spaces of Figs. \ref{fig:commm2_dr}, \ref{fig:commm4_dr} and \ref{fig:commm6_dr} of the paper and Figs. S1-S9 in the Supplemental Material \cite{SM} corresponds to a single realization of the system described in Subsec. \ref{Subsection_Hindmarsh-Rose_Model_for_Brain_Dynamics}.

\subsection{Stereotypical Electrical Stimuli}\label{Subsection_stereotypical_electrical_stimuli}

Sensory organs receive stimuli which are represented by trains of short electrical impulses. These excitations can be thought of as stereotypical electrical events appearing in the form of spikes. As an example we refer to the stimuli that arrive in the auditory system. The stimuli are converted by the cochlea into trains of spikes which are received by groups of neurons tuned to different frequencies \cite{PhysRevLett.91.128101}. The amplitudes of the spikes by themselves are not important, but what is rather important is the strength of the stimuli which is encoded in the interspike intervals, or in the number of spikes per time unit (firing rate). Electrical impulses produce a current density that can be regarded as an input current $I^{\mbox{ext}}$.

In this work, we excite single communities of the \textit{C.elegans} BDN independently by supplying external electrical stimuli, following a Poisson process, to all neurons of the perturbed community, i.e. the histogram of interspike intervals follows a Poisson distribution. We then repeat this process for all communities, perturbing them one by one, keeping the other five in an unperturbed state. Poisson processes are characterized by the fact that the probability of a neuron to fire a spike is independent of its past firing activity, and thus the interspike intervals are independent of the spiking history. We denote by $\tau_k=t_k-t_{k-1}$ the $k$-th interspike interval and use the Poisson process \cite{2014PhyA..410..628B}
\begin{equation}\label{poisson_process}
P(\tau)=1-e^{-r\tau},
\end{equation}
where $r$ is the input spiking rate and $P(\tau)dt$ the probability of finding an interspike interval between $\tau$ and $(\tau+d\tau)$. In our study, we use a sequence of stereotypical impulses that correspond to a maximum external current $I^{\mbox{ext}}=3.25\mu\mbox{A/cm}^2$ so that the interspike intervals $\tau_k$ satisfy the Poisson process \eqref{poisson_process} with a given input rate $r$. For the neurons of the unperturbed communities, we use the constant external current $I^{\mbox{ext}}=1.3\mu\mbox{A}/\mbox{cm}^2$ to allow them to function initially in a low-neural activity state. For the perturbed communities, the external electrical stimulations start after the transient time interval $t_t=300$ and continue until $t_f=5000$. We present such an example in Fig. \ref{fig:histogram}.

\subsection{Estimation of the Dynamic Range and Steven Law Exponent}\label{Subsection_estimation_dynamic_range}

Following the approach in Ref. \cite{2014PhyA..410..628B}, for each pair of chemical and electrical couplings in the parameter spaces of Figs. \ref{fig:commm2_dr}, \ref{fig:commm4_dr}, \ref{fig:commm6_dr} and for all (S1-S9) in the Supplemental Material \cite{SM}, we estimate the dynamic range $\Delta$ as follows: As observed in Fig. \ref{fig:average_firing_rate_p}, there is typically a range of input rates $r$ for which the firing rate $F$ exhibits a sigmoidal behavior, known as Steven law \cite{Stevens1975}. In particular, $F\sim r^m$, where $m$ is Steven exponent. Since the bounds for $r$, for the estimation of the slope of the sigmoidal curve are not fixed as they depend on the particular neural dynamics and responses to external perturbations $I^{\mbox{ext}}$, one can consider the bounds $F_{\mbox{low}}=0.2(F_{\mbox{max}}-F_{\mbox{min}})$ and $F_{\mbox{upp}}=0.9(F_{\mbox{max}}-F_{\mbox{min}})$ to compute $m$. Then, the dynamic range is given by
\begin{equation}
\Delta=10\log_{10}\biggl(\frac{r_{\mbox{upp}}}{r_{\mbox{low}}}\biggr),
\end{equation}
measured in decibels (dB). $r_{\mbox{low}}$ and $r_{\mbox{upp}}$ correspond to $F_{\mbox{low}}$ and $F_{\mbox{upp}}$ respectively, based on the $F$ vs $r$ plot. The exponent $r$ can be calculated by fitting the data in the interval [$r_{\mbox{low}},r_{\mbox{upp}}]$ with the power-law $F(r)\propto r^m$.

\subsection{Community Detection Methodology}\label{Subsection_analysis_of_communities}

\begin{figure}[!ht]
\centering{
\includegraphics[scale=0.85]{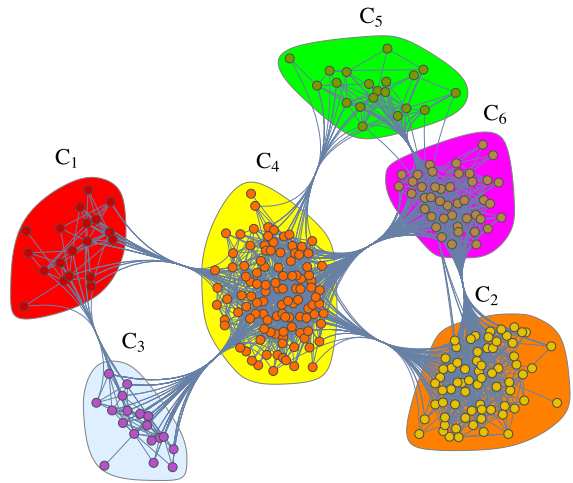}
}
\caption{\textbf{The six communities identified by the walktrap community detection method in the \textit{C.elegans} brain network, based on \cite{Antonopoulosetal2015}}. Each color region represents a single community denoted as C$_i$, $i=1,2,\ldots,6$. Note that the communities have different sizes.}\label{fig:walktrap_communities_full_graph}
\end{figure}

We use in our study the same communities and, corresponding electrical and chemical adjacency matrices reported in \cite{Antonopoulosetal2015}. The authors there used the walktrap community detection method \cite{Ponsetal2005} with six steps to identify them. The algorithm detects communities through a series of short random walks, with the idea that the vertices encountered on any given random walk are more likely to be within a community. It initially treats all nodes as communities of their own, then merges them into larger, and these into still larger, and so on. Essentially, it tries to find densely connected subgraphs (i.e. communities) in a graph via random walks. The idea is that short random walks tend to stay in the same community. This approach was able to identify 6 communities of different sizes in the \textit{C.elegans} brain network, plotted here in Fig. \ref{fig:walktrap_communities_full_graph}.

\section{Results}\label{section_results}

\subsection{Stimulus-Response Relation in the \textit{C.elegans} BDN}

Neurons in the human brain are connected with themselves and form a complex network which consists of about 86 billion neurons on average, with each neuron connected to about $10^4$ others, yielding about $10^{15}$ synapses in total\cite{Azevedoetal2009}. Macroscopically, and to simplify its structural complexity, one can describe this complex network by considering cortical areas as its nodes, connected by axonal fibres. This can be justified by the fact that cortical areas consist of ensembles of neurons that perform the same or similar functions and, thus, may act as individual units. This is happening for example when the brain is performing a particular task or when the neural ensembles process the same kind of external stimuli \cite{Scannell1993191}.

In this work however we consider the brain network of the connectome of the large somatic nervous system of the \textit{C.elegans} soil worm which is almost completely mapped and consists of 277 neurons and about 7000 synapses \cite{cmtkdataset} (see Subsec. \ref{subSubsection_C.elegans_data} of the Methodology for more details). We do so to avoid the complexity of larger brain networks, such as those derived from human subjects for example, without loosing similar structural properties, i.e. the small-world and modular structure. In particular, we consider the undirected version of the connectivity matrix provided in Ref. \cite{cmtkdataset}, resulting in a symmetric adjacency matrix. In Ref. \cite{Antonopoulosetal2015}, the authors performed a community detection analysis for this brain network using the walktrap method (see Subsec. \ref{Subsection_analysis_of_communities} of the Methodology for more details) and identified groups of neurons with similar topological characteristics that were characterized as members of the same community. This approach partitions the \textit{C.elegans} brain network into six, interconnected, small-world communities with different sizes (see Fig. \ref{fig:walktrap_communities_full_graph}), and is reminiscent of findings in neuroscience about the modular organization of the brain\cite{Meunieretal2001}. It is also in agreement with findings in Refs. \cite{Wattsetal1998,Antonopoulosetal2015} that suggest that the nervous system of the nematode \textit{C.elegans} worm has a small-world structure.

Here, we aim to study systematic perturbations in the communities of the \textit{C.elegans} brain network and their response to external stimuli. The study is inspired by a more realistic neuroscientific experiment where one provides external electrical stimuli to one part of the brain and studies how and if the stimulations trigger persistent neural activity in other cortical regions. This is related to the so-called binding problem, i.e. how items encoded by distinct brain circuits can be combined for the development of perception, decision-making, and action involving other cortical areas \cite{Feldman2012}. In our approach, we perturb the neurons of a single community by providing trains of stereotypical electrical Poissonian impulses and by constantly increasing the input rate $r$. These stereotypical impulses take the form of train spikes forming the external current $I^{\mbox{ext}}$ for the perturbed ones. In this framework, we keep the other five communities unperturbed, allowing them to function at a base level of fixed $I^{\mbox{ext}}=1.3\mu\mbox{A}/\mbox{cm}^2$, emulating neural activity in the absence of perturbations, characterized by persisting spiking activity with no quiescent periods. Then, we study if the propagation of the initial stimulations in one community trigger neural activity to the others by measuring the dynamic range $\Delta$ and Steven law exponent $m$ for all communities and for the full \textit{C.elegans} brain network. We explain the technical details of our study in Subsec. \ref{Subsection_estimation_dynamic_range} of Methodology.

\begin{figure}[!ht]
\centering{
\includegraphics[scale=0.62]{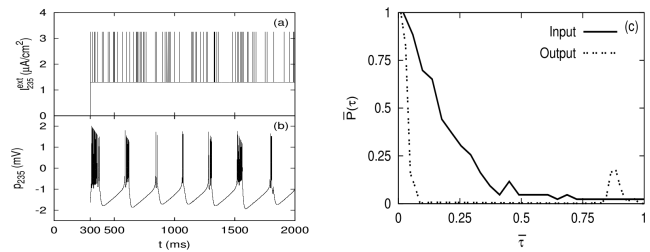}
}
\caption{\textbf{A Poissonian electrical stimulation (Input), its corresponding neural response (Output) and histogram of interspike intervals.} In panel (a), taken from Fig. \ref{fig:commm2_dr}, the input electrical stimulation $I^{\mbox{ext}}_{235}$ that follows the Poisson process of Eq. \eqref{poisson_process} with $r=0.05$, excites neuron 235 of the second community and, in panel (b), the corresponding spike activity $p_{235}$ of the same neuron. Panel (c) shows the normalized histogram $\bar{\mathcal{P}}(\tau)$ of the normalized interspike interval $\bar{\tau}$ for the external stimulus $I^{\mbox{ext}}_{235}$ (black solid curve) and for the spike activity $p_{235}$ (dashed dotted black curve) of the previous two panels. Quantities in both axes of panel (c) are normalized with respect to their maximum values. The neurons of the second community are electrically stimulated during the time interval $[300,5000]$.}\label{fig:histogram}
\end{figure}

We present in Fig. \ref{fig:histogram}a) a typical example of Poissonian electrical impulses in the form of train spikes that form the external current $I^{\mbox{ext}}_{235}$ acting on neuron 235 of the second community. In this example, we set $r=0.05$. Similar external currents are also applied to all neurons of the second community to implement an external perturbation that takes place in only one part of the BDN, i.e. in the second community. The response of neuron 235 under such a perturbation is shown in panel (b). Due to the intrinsic refractory period of the neurons, not all electrical perturbations (spikes) are able to trigger new action potentials, and thus neurons remain quiescent during this time period. The normalized histogram $\bar{\mathcal{P}}(\tau)$ of the normalized interspike interval $\bar{\tau}$ for the external stimulus $I^{\mbox{ext}}_{235}$ (black solid curve) of panel (a) and for the spike activity $p_{235}$ (dashed dotted black curve) of panel (b) are shown in panel (c), where ``Input'' denotes the applied external stimulus and ``Output'' the neural response. Both histograms for the input and output activity closely follow a Poissonian distribution as the interspike intervals are the result of a Poissonian process.

\begin{figure}[!ht]
\centering{
\includegraphics[scale=1]{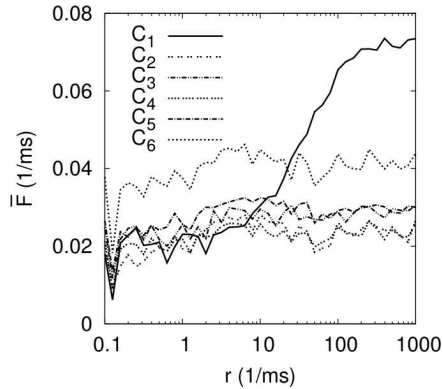}
}
\caption{\textbf{Example of stereotypical stimulus-response behaviors for HR neurons in the \textit{C.elegans} BDN.} Stereotypical behavior of the average firing rate $\bar{F}$ based on the $p$ variable of the HR system for the six communities (C$_1$ to C$_6$) as the input rate $r$ of the Poisson process increases. In this example, we have perturbed the neurons of the first community, similarly to panel (a) of Fig. \ref{fig:histogram}.}\label{fig:average_firing_rate_p}
\end{figure}

The firing rate $F$ of a single neuron characterizes its response to a stimulus $I^{\mbox{ext}}$. $F$ can be estimated by varying the input rate $r$ of the Poisson process that corresponds to the interspike intervals of the external stimulus. Typically, cortical neurons receive sensory input ranging from $10^4$ to $10^5$ spikes per second that corresponds to input rates $r=10$ to 100(ms)$^{-1}$, depending on the case \cite{RIS_0,2014PhyA..410..628B}. However, in peripheral organs this rate is substantially slower, of the order of 0.1 to 1(ms)$^{-1}$ due to larger conductance changes \cite{Gerstneretal2002}. In our work, we decided to extend the range of input rate $r$ from 0.1 to 1000(ms)$^{-1}$ to capture all possible neural response effects as well as the saturation of the stimulus-response curves. Such an example can be seen in Fig. \ref{fig:average_firing_rate_p}.

Particularly, in this figure we present the behavior of the average firing rate $\bar{F}$ of the first community (denoted as C$_1$ in the figure) as the input rate $r$ of the external stimulation, applied to the neurons of the first community, is increased from 0.1 to 1000(ms)$^{-1}$. $\bar{F}$ is the average firing rate with respect to the firing rates of all neurons of the community. A sigmoidal relation can be seen for $\bar{F}$ and $r$ that characterizes the collective neural response of the first community (black solid curve) but not observed for the other five communities, for which the stimulus-response curves appear quite flat and not sigmoidal-shaped (see C$_2$-C$_6$ communities in Fig. \ref{fig:average_firing_rate_p}). These results already demonstrate that external perturbations in the first community are unable to trigger similar responses to the other five, which essentially remain quite insensitive. In Fig. S1 of the Supplemental Material \cite{SM}, we show that the dynamic range $\Delta$ of the other five communities is actually very small compared to the first one, for all chemical and electrical coupling pairs considered in the parameter space. This is further evidenced in Fig. S4 of the Supplemental Material \cite{SM} where we show that the Steven law exponent $m$ of all but the first community, are quite small, implying that the stimulus-response curves are indeed flat.

Typically, there are three regimes with respect to $r$ in the case the stimulus-response curve is sigmoidal. For example, in Fig. \ref{fig:average_firing_rate_p}, for very small $r$ between 1 to 10(ms)$^{-1}$, there is a small increase of the neural response to the external perturbations. The second occurs for $r$ values in 10 to 100(ms)$^{-1}$ in which the average firing rate of the first community steadily increases (see the solid black curve) until it reaches its maximal value. The third one corresponds to the saturation regime that happens in this case for $r>100$(ms)$^{-1}$. In the first regime, the average neural response of the first community increases very little, since spikes follow closely the input signal and $r$ is small, so is the spike rate of the neurons, leading to small average firing rates. In the second regime, the firing rates are able to follow the increase of the input rate $r$. However, this situation changes drastically in the third regime of large $r$ (i.e. $r>100$(ms)$^{-1}$), in which the refractory periods of the neurons are responsible for the slow down or constancy of the average firing rate, i.e. the saturation regime. This is due to the refractory periods of the neurons, during which they remain insensitive to external perturbations.

\begin{figure}[!ht]
\centering{
\includegraphics[scale=0.66]{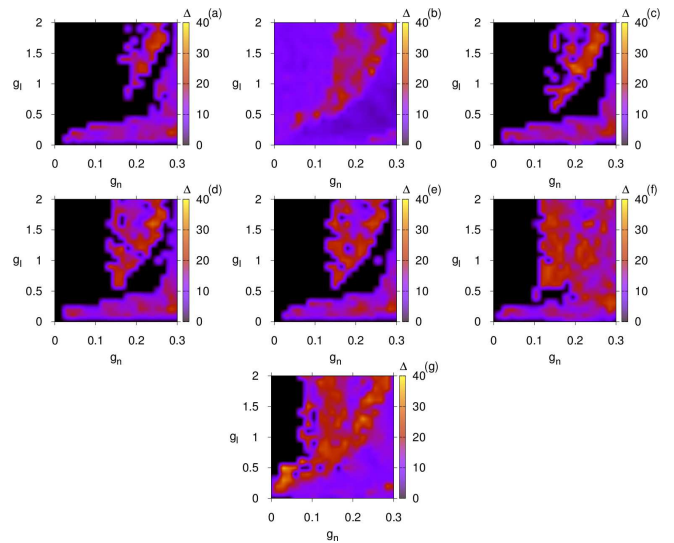}
}
\caption{\textbf{The dynamic range $\boldsymbol{\Delta}$ for the six communities of the \textit{C.elegans} BDN when perturbing the second community.} Parameter spaces for communities 1 (panel (a)) to 6 (panel (f)) for the dynamic range $\Delta$. Panel (g) is the parameter space of the dynamic range $\Delta$ computed for the whole BDN. All dynamic ranges are expressed in dB.}\label{fig:commm2_dr}
\end{figure}

So far, we have shown that external perturbations in the first community do not trigger similar firing rate responses to the neurons of the other five communities and that they essentially remain insensitive to external perturbations. We extend this result and show in Figs. S2 and S3 of the Supplemental Material \cite{SM} that this is also happening for the third and fifth communities. In the following, we analyze if this is a generic behavior exhibited by all communities in the \textit{C.elegans} BDN or if, depending on which one is perturbed, interesting new phenomena will emerge with respect to which ones are consequently activated and deactivated, as measured by their corresponding dynamic ranges.
We have thus performed a similar analysis for the stimulus-response behavior for the other communities and found that when we perturbed the second, fourth and sixth communities, the external stimulations were able to propagate to some of them but not to all and that they triggered neural responses similar to those of the initially perturbed one. We note here that the perturbations did not propagate to all of them. We show in Fig. \ref{fig:commm2_dr} such an example in which  external perturbations applied to the second community (panel (b)) trigger similar firing rate responses mainly to the neurons of the fourth (panel (d)), fifth  (panel (e)) and sixth (panel (f)) and less, to the first (panel (a)) and second (panel (b)) communities. This is also evident in panel (g) for the dynamic range computed for the whole BDN. In Fig. S5 of the Supplemental Material \cite{SM} we present the corresponding parameter spaces for the Steven law exponents $m$ for all communities and the full BDN and, observe that in regions where $\Delta$ is large, $m$ is relatively smaller than in other regions with much smaller $\Delta$. This is expected since large dynamic ranges occur when the average firing rates constantly increase for large enough input rate intervals, giving rise to relatively small Steven law exponents $m$, as $m$ is the slope of the main, linearly increasing, part of the stimulus-response curve. We present similar results in Figs. \ref{fig:commm4_dr}, \ref{fig:commm6_dr} and, S7, S9 of the Supplemental Material \cite{SM} for external perturbations initially applied to the fourth and sixth communities, with possibly slight variations with respect to the amount of the collective neural activation, such as in the first community in Fig. \ref{fig:commm6_dr}.

Our results for the large dynamic ranges (i.e. the red and orange regions in the parameter spaces of Figs. \ref{fig:commm2_dr}, \ref{fig:commm4_dr} and \ref{fig:commm6_dr}) complement those reported in Ref. \cite{Antonopoulosetal2015} (see Fig. 2B in there). Particularly, the authors found that the coupling strength regions for which the information flow capacity $I_c$ is maximal in the \textit{C.elegans} BDN (red region in Fig. 2B in Ref. \cite{Antonopoulosetal2015}) are similar to those for high dynamic ranges identified in this work. Since, maximal $I_c$ is attributed to low neural synchronization and self-criticality \cite{Antonopoulosetal2015} in this region, we associate the large dynamic ranges obtained here to these two important properties exhibited by the BDN. In other words, self-criticality, low level neural synchronization and maximal information flow capacity seem to be prominent conditions for the \textit{C.elegans} BDN to respond to external electrical perturbations. What is even more fascinating and reminiscent of realistic neurophysiological experiments is that these responses are triggered when stimulating specific communities of the BDN and that it does not happen for all of them. A possible interpretation of these findings is that when the BDN is responding to external perturbations, the communities are able to exchange the largest possible amounts of information between them, i.e. maximize their information flow capacity.

\begin{figure}[!ht]
\centering{
\includegraphics[scale=0.66]{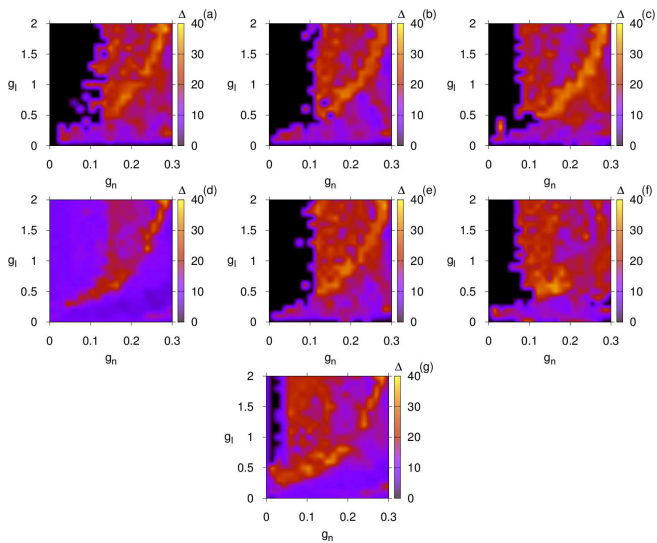}
}
\caption{\textbf{The dynamic range $\boldsymbol{\Delta}$ for the six communities of the \textit{C.elegans} BDN when perturbing the fourth community.} Parameter spaces for communities 1 (panel (a)) to 6 (panel (f)) for the dynamic range $\Delta$. Panel (g) is the parameter space of the dynamic range $\Delta$ computed for the whole BDN. All dynamic ranges are  expressed in dB.}\label{fig:commm4_dr}
\end{figure}

\begin{figure}[!ht]
\centering{
\includegraphics[scale=0.66]{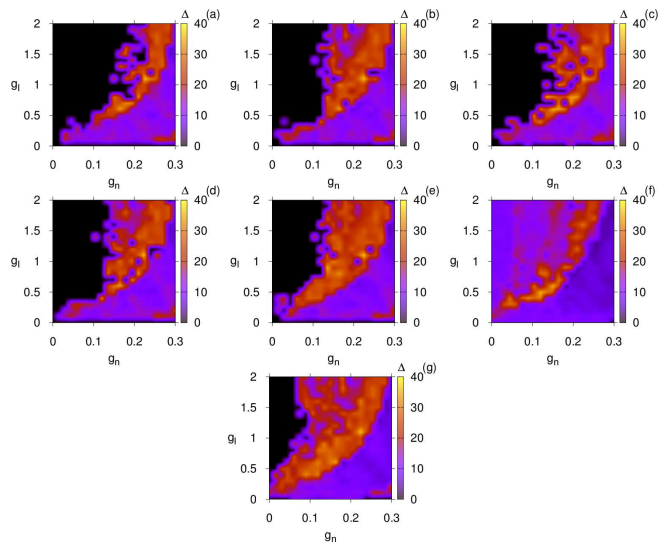}
}
\caption{\textbf{The dynamic range $\boldsymbol{\Delta}$ for the six communities of the \textit{C.elegans} BDN when perturbing the sixth community.} Parameter spaces for communities 1 (panel (a)) to 6 (panel (f)) for the dynamic range $\Delta$. Panel (g) is the parameter space of the dynamic range $\Delta$ computed for the whole BDN. All dynamic ranges are expressed in dB.}\label{fig:commm6_dr}
\end{figure}

\section{Conclusions}\label{section_conclusions}

In this paper we studied the effect of the application of external electrical stimuli to specific communities of the brain dynamical network of the \textit{C.elegans} soil worm, given by the connectome of its large somatic nervous system. We systematically studied external, electric perturbations and their responses in its brain dynamic network, and modelled the external stimulations by Poisson processes of fixed average input rate. We aimed to study how these perturbations spread to other communities and for which coupling strengths the spreading occurs.

The partition of the \textit{C.elegans} brain network into communities is reminiscent of the modular organization of the brain. Our study was inspired by an experiment in which external electrical stimuli are applied to a specific part of the brain to see how and if persistent neural activity is triggered in other cortical regions. We showed that when one perturbs specific communities such as the second, fourth and sixth, keeping the rest unperturbed, the external stimulations are able to propagate to other communities of the brain network. That depends on the initially perturbed one. Our approach allowed us to understand for which coupling regions and communities, the external perturbations spread to others by computing the power-law dependence of the stimulus-response relation and by estimating the dynamic range and Steven law exponent.

Our results complement those reported recently in Ref. \cite{Antonopoulosetal2015}, as the coupling strength regions for which the information flow capacity was found to be maximal in the \textit{C.elegans} brain dynamical network\cite{Antonopoulosetal2015} are similar to those found here for high dynamic ranges. In Ref. \cite{Antonopoulosetal2015}, information flow capacity was found to become maximal when neural synchronization is low, for coupling strengths for which the system works at self-criticality with only one positive Lyapunov exponent. The latter behaviors are associated with the large dynamic ranges found here as they occur for the same coupling strength regions, and imply that self-criticality, low level neural synchronization and maximal information flow capacity seem to be prominent conditions for the \textit{C.elegans} brain dynamical network to respond to external electrical perturbations. In this case, the responses are triggered when stimulating specific communities of the brain dynamical network and not just any. We interpreted this behavior as the ability of the brain network to respond to external perturbations when it works at self-criticality and its information flow capacity becomes maximal. Finally, we believe it would be interesting to extend this study in brain networks coming for example from human subjects to check if self-criticality, maximal information flow and large dynamic ranges are also related to the ability of these brain networks to respond to external perturbations, similarly to what was found here for the \textit{C.elegans} brain network.



%
%

%

\begin{acknowledgments}
We are grateful for fruitful discussions with F. S. Borges and E. L. Lameu.
\end{acknowledgments}


\providecommand{\noopsort}[1]{}\providecommand{\singleletter}[1]{#1}%

\end{document}